\documentclass[twocolumn,linenumber,trackchanges]{aastex631}

\usepackage{savesym}
\savesymbol{tablenum}
\usepackage[group-digits=false,separate-uncertainty]{siunitx}
\restoresymbol{SIX}{tablenum}
\begin{document}
\title{ALMA reveals spatially-resolved properties of molecular gas in the host galaxy of FRB\,20191001A at z = 0.2340}
\author[0009-0001-4699-5811]{Itsuki Yamanaka}
\affiliation{Institute of Astronomy, Graduate School of Science, The University of Tokyo, 2-21-1 Osawa, Mitaka, Tokyo 181-0015, Japan}
\author[0000-0001-6469-8725]{Bunyo Hatsukade}
\affiliation{National Astronomical Observatory of Japan, 2-21-1 Osawa, Mitaka, Tokyo 181-8588, Japan}
\affiliation{Graduate Institute for Advanced Studies, SOKENDAI, Osawa, Mitaka, Tokyo 181-8588, Japan}
\affiliation{Institute of Astronomy, Graduate School of Science, The University of Tokyo, 2-21-1 Osawa, Mitaka, Tokyo 181-0015, Japan}
\author[0000-0002-1639-1515]{Fumi Egusa}
\affiliation{Institute of Astronomy, Graduate School of Science, The University of Tokyo, 2-21-1 Osawa, Mitaka, Tokyo 181-0015, Japan}
\author[0000-0001-7228-1428]{Tetsuya Hashimoto}
\affiliation{Department of Physics, National Chung Hsing University, No. 145, Xingda Rd., South Dist., Taichung, 40227, Taiwan (R.O.C.)}
\author[0000-0001-5322-5076]{Yuu Niino}
\affiliation{Kiso Observatory, Institute of Astronomy, Graduate School of Science, The University of Tokyo, 10762-30 Mitake, Kiso-machi, Kiso-gun, Nagano 397-0101, Japan}
\author[0000-0002-0944-5634]{Tzu-Yin Hsu}
\affiliation{Department of Physics, National Tsing Hua University, 101, Section 2. Kuang-Fu Road, Hsinchu, 30013, Taiwan (R.O.C.)}
\author[0000-0002-2699-4862]{Hiroyuki Kaneko}
\affiliation{Department of Environmental Science, Faculty of Science,
8050 Ikarashi 2-no-cho, Nishi-ku, Niigata, Niigata 950-2181, Japan}
\author[0000-0002-4052-2394]{Kotaro Kohno}
\affiliation{Institute of Astronomy, Graduate School of Science, The University of Tokyo, 2-21-1 Osawa, Mitaka, Tokyo 181-0015, Japan}

\begin{abstract}
We report the detection of the CO(2--1) emission line with a spatial resolution of $0\farcs9$ ($\SI{3.5}{kpc}$) from the host galaxy of the fast radio burst (FRB), FRB\,20191001A at $z=0.2340$, using the Atacama Large Millimeter/submillimeter Array. 
This is the first detection of spatially resolved CO emission from the host galaxy of an FRB at a cosmological distance.
The inferred molecular gas mass of the host galaxy is $\SI{2.3+-0.4e+10}{M_\odot}$, indicating that it is gas-rich, as evidenced by the measured molecular gas fraction $\mu_\mathrm{gas}=\num{0.50+-0.22}$.
This molecular-gas mass and the star formation rate of the host, $\mathrm{SFR}=\SI{8.06+-2.42}{M_\odot. yr^{-1}}$, differ from those observed in the other FRB host galaxies with the average $M_\mathrm{gas}=\SI{9.6e+8}{M_\odot}$ and $\mathrm{SFR}=\SI{0.90}{M_\odot. yr^{-1}}$.
This lends further credibility to the hypothesis that FRBs may originate from single or multiple progenitors across a diverse range of galaxy environments.
Based on the observed velocity field modeling, we find that the molecular gas disk is dominated by an ordered circular rotation, despite the fact that the host galaxy has a gas-rich companion galaxy with a projected separation of $\sim 25$ kpc. 
The formation of the FRB's progenitor might not have been triggered by this interaction.
We derive the 3$\sigma$ upper limit of the molecular gas column density at the FRB detection site to be $< 2.1\times 10^{21} \,\mathrm{cm^{-2}}$ with a 3$\sigma$ upper limit.
\end{abstract}

\keywords{
\href{http://astrothesaurus.org/uat/2008}{Radio transient sources (2008)};
\href{http://astrothesaurus.org/uat/1339}{Radio bursts (1339)};
\href{http://astrothesaurus.org/uat/508}{Extragalactic radio sources (508)};
\href{http://astrothesaurus.org/uat/602}{Galaxy kinematics (602)};
\href{http://astrothesaurus.org/uat/262}{CO line emission (262)};
\href{http://astrothesaurus.org/uat/1073}{Molecular gas (1073)};
\href{http://astrothesaurus.org/uat/847}{Interstellar medium (847)};
}

\section{Introduction}
Fast Radio Bursts (FRBs) are bright radio pulses observed with dispersed sweep, first reported by \cite{Lorimer2007}. 
The pulse timescale ranges from microseconds to milliseconds, and their dispersion measure and the localization of FRB \citep[e.g.][]{Nimmo2022} suggest that they have an extragalactic origin \citep[e.g.][]{Thornton2013}. 
Although many models have been proposed, such as compact object mergers, magnetars, neutron stars, supernovae (SNe), and gamma-ray bursts (GRBs) \citep{Platts2019}, there is no definitive explanation for the origin of FRBs, making it one of the most important questions in modern astronomy. 
Given the expected correlation between the occurrence environment of FRBs and the galactic environment, recent approaches to understanding the origin of FRBs involve observing the host galaxies. 
So far, about 30 FRB host galaxies have been identified, and optical/near-infrared studies have shown a wide range of stellar masses and star formation rates \citep{Heintz2020, Gordon2023}. 
To further understand star formation in galaxies, observing molecular gas, the material for star formation, is an effective approach. 
Since a hydrogen molecule does not emit electromagnetic waves at the temperatures typical of molecular clouds, it is standard to use CO lines to derive molecular gas properties.
The CO observations of FRB host galaxies by \cite{Hatsukade2022} showed that the molecular gas mass ranges widely ($\sim 2.5$ orders of magnitude). 
\citet{Chittidi2023} showed that FRB host galaxies belong to a different population than that of star-forming galaxies using the Kaplan-Meier estimator for molecular gas fractions $\mu_\mathrm{gas}$. 
However, these studies do not provide statistical suggestions on the specific origin of FRBs due to the small sample size (9 galaxies for CO observations, but 4 of them are non-detected), so further CO observations of FRB host galaxies are needed.

Observations of molecular gas (and HI gas) can reveal the interaction between galaxies, such as gas inflows, outflows, and mergers. 
HI observations have shown gas turbulence and mergers \citep{Kaur2022}, and asymmetric motion spectra \citep{Micha2021,Glowacki2023} in FRB host galaxies.
CO spectra of FRB host galaxies are also found to be asymmetric \citep{Hsu2022}.
These studies suggest that recent star formation activity was triggered by galaxy interactions, which may have led to FRB progenitor activity. 
However it is not clear if these emission-line profiles are result of interaction because the galaxies are not spatially resolved.
Therefore, it is needed to resolve FRB host galaxies to find whether galactic interaction is really a critical condition for FRBs.

Furthermore, to identify the direct environment of FRBs, it is essential to determine the molecular gas properties at the FRB location. 
In this study, we report the observations of the host galaxy of FRB\,20191001A and its companion galaxy, which is thought to interact with the FRB host, using the Atacama Large Millimeter/submillimeter Array (ALMA) with a spatial resolution of $\SI{3.5}{kpc}$. 
We derive molecular gas properties of these galaxies and at the FRB location and discuss its progenitors and possible effects of the galactic interaction.

Throughout this paper, we adopt cosmological parameters $H_0=\SI{69.6}{km.s^{-1}.Mpc^{-1}}$ and $\Omega_\mathrm{M}=0.286$ from \cite{Bennett2014} to calculate physical values.

\section{Observations}
\subsection{Target}
FRB\,20191001A was detected by the Australian Square Kilometre Array Pathfinder (ASKAP) on 2019 October 1 at 16:55:35.97081 UT, and localized to a host galaxy DESJ213324.44-544454.65 at $z=0.2340$ (hereafter HG\,20191001A, or simply ``the host'') \citep{Bhandari2020}. 
The stellar mass is $M_\star=\SI{4.6+-1.9e+10}{M_\odot}$, star formation rate (SFR) is $\SI{8.1+-2.4} {M_\odot. yr^{-1}}$, and gas-phase metallicity is $12+\log(\mathrm{O/H})=\num{8.94+-0.05}$ \citep{Heintz2020}.
It is located at the massive end of the star formation main sequence on the $M_\star - \mathrm{SFR}$ diagram. 
The galaxy J213323.65-544453.6 at $z=0.2339$ (hereafter ``the western source'') is located $\sim\SI{25}{kpc}$ to the west of this host and is considered to be in a physical interaction (Figure \ref{I-band}). 
The SFR of the western source is estimated to be about $\SI{21}{M_\odot.yr^{-1}}$ based on the 1.4 GHz luminosity \citep{Bhandari2020}.
Properties of the FRB, its host, and the western source are summarized in Table \ref{FRBs}.

\begin{table}[htbp]
  \caption{Properties of the FRB\,20191001A, its host galaxy, and the western source}
  \label{FRBs}
  \centering
    \begin{tabular}{lcc}
      \hline \hline
      Parameter & Value & Ref.\\
      \hline
      \uline{FRB\,20191001A}&&\\
       $\mathrm{R.A.}$ (J2000)&$\mathrm{21^h33^m24.373^s}$&1\\
       $\mathrm{Decl.}$ (J2000) &$\mathrm{-54^\circ 44\arcmin 51.43\arcmin\arcmin}$&1\\
       DM ($\si{pc.cm^{-3}}$)&506.92&1\\
      \hline
      \uline{HG\,20191001A}&&\\
      $\mathrm{R.A.}$ (J2000)&$\mathrm{21^h33^m24.44^s}$&2\\
      $\mathrm{Decl.}$ (J2000)&$\mathrm{-54^\circ 44\arcmin 54.65\arcmin\arcmin}$&2\\
      $z$&0.2340&2\\
      $\mathrm{SFR_{H\alpha}} (M_\odot\,\mathrm{yr^{-1}})^{*}$&$\num{8.06+-2.42}$&2\\
      $\mathrm{SFR_{Radio}} (M_\odot\,\mathrm{yr^{-1}})^{**}$&$11.2$&1\\
      $M_\star$($M_\odot$)&$\num{4.64+-1.88e+10}$&2\\
      $Z=12+\log(\mathrm{O/H})$&$\num{8.94+-0.05}$&2\\
      \hline
      \uline{the western source}&&\\
      $\mathrm{R.A.}$ (J2000)&$\mathrm{21^h33^m23.65^s}$&2\\
      $\mathrm{Decl.}$ (J2000)&$\mathrm{-54^\circ 44\arcmin 53.6\arcmin\arcmin}$&2\\
      $z$&0.2339&2\\
      $\mathrm{SFR_{Radio}} (M_\odot\,\mathrm{yr^{-1}})^{**}$&$20.6$&1\\
      \hline
    \end{tabular}
    \tablecomments{$^{*}$SFR estimated from H$\alpha$ line luminosity.\, $^{**}$SFR estimated from 1.4 GHz luminosity. Ref. 1: \cite{Bhandari2020}, 2: \cite{Heintz2020}}
\end{table}

\subsection{ALMA observations and data reduction}
A CO(2--1) observation of FRB\,20191001A were made on 2022 September 25, from 01:15:40 to 01:49:37 UTC with ALMA Band 5 (Project code is 2021.1.00027.S).
45 antennas were set up with baseline lengths of 15.1--500.2 m. The total integration time on the science target was 11 min 9 sec. 
The spectral windows are 25, 27, 29, 31, and the correlator setup bandwidth was 1.875 GHz and subdivided into 1920 channels for each spectral window.

Data reduction was performed with the Common Astronomy Software Application (CASA; \citealt{TheCASATeam_2022}), with the CASA pipeline version 6.4.1.12. 
Imaging was performed using the tclean task, with a cell size of $0\farcs2$, a weighting of briggs, a robust parameter of 2, a threshold of twice the rms noise, and used a mask of CO emission area. 
The data cube was generated over the velocity range from $\SI{-1000}{km.s^{-1}}$ to $\SI{1000}{km.s^{-1}}$ with a velocity resolution of $\SI{50}{km.s^{-1}}$.
In the cube, $\SI{0}{km.s^{-1}}$ is defined from the restframe frequency calculated from the host's optical redshift, which is based on observed wavelengths of H$\beta$, [\ion{O}{3}]$\lambda 5007$, H$\alpha$, and [\ion{N}{2}]$\lambda 6583$ lines.
The continuum map was created by excluding channels with CO emission. 
The rms noise level was $\SI{0.92}{mJy.beam^{-1}}$ for each channel of the $\SI{50}{km.s^{-1}}$ cube and $\SI{3.0e-2}{mJy.beam^{-1}}$ for the continuum map.
The synthesized beam sizes were $3.5\times3.4\,\si{kpc^2}(0\farcs94 \times 0\farcs91)$ and $4.1\times3.8\si{kpc^2}(1\farcs1 \times 1\farcs0)$, respectively.

\section{Results}
\begin{figure}[htbp]
    \centering
    \includegraphics[width=8.5cm]{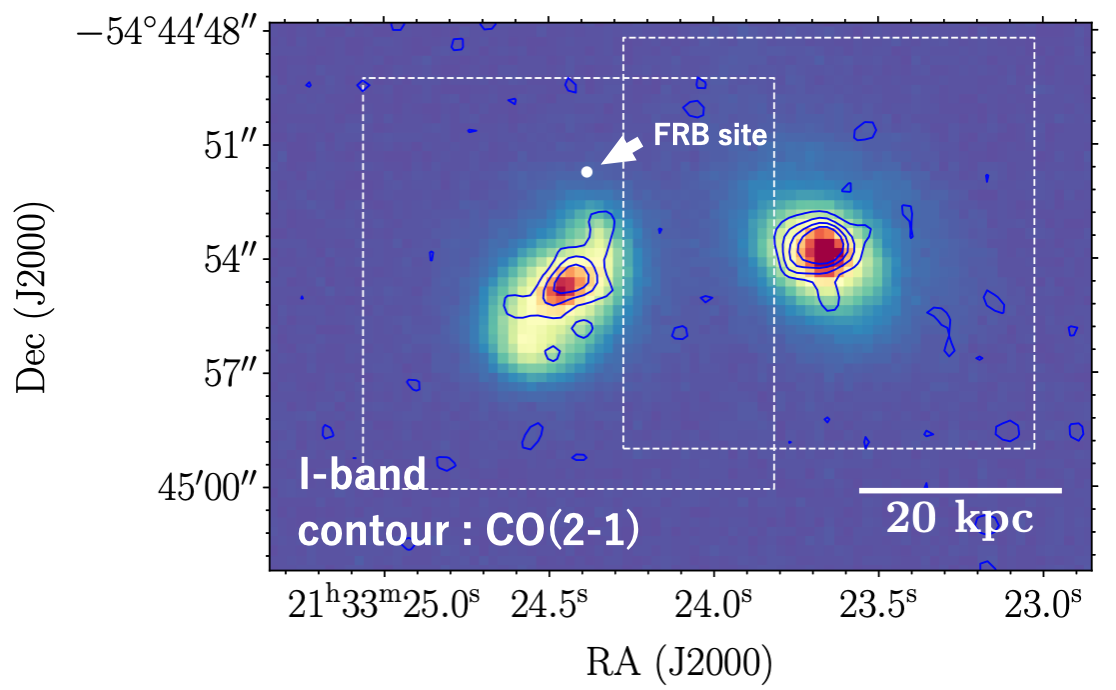}
    \caption{The \textit{I}-band image of HG\,20191001A and the western source. The observed CO(2--1) velocity-integrated intensity images are overplotted as blue contours (2$\sigma$, 4$\sigma$, 6$\sigma$, and 8$\sigma$ where $1\sigma = \SI{0.30}{Jy. beam^{-1}. km. s^{-1}}$). The white filled circle shows the location where FRB\,20191001A is detected and its detection error range \citep{Bhandari2020}. Two white dashed rectangles correspond to the areas of Figure \ref{obsresults}.}
    \label{I-band}
\end{figure}

The CO(2--1) integrated intensity map is created by summing all the channels in the cube with no threshold and presented as blue contours in Figure \ref{I-band}. The results for each galaxies are also shown in Figure \ref{obsresults}.
CO(2--1) emission line was detected in both galaxies, and the spectra were extracted from the area containing each galaxy.
The results of the analysis in this section are shown in Table \ref{results}.

\begin{table}[htbp]
  \caption{Results of the observation and analysis}
  \label{results}
  \centering
    \begin{tabular}{lc}
      \hline \hline
      Parameter & Value \\
      \hline
      \uline{FRB\,20191001A}&\\
      $N(\mathrm{H_2})$($\si{cm^{-2}}$)$^{*}$&$< \num{2.1e+21}$\\
      $S_\mathrm{1.3mm}$($\si{mJy.beam^{-1}}$)$^{*}$&$< \num{9.1e-2}$\\
      \hline
      \uline{HG\,20191001A}&\\
      $S_\mathrm{CO}\Delta V$($\si{Jy.km.s^{-1}}$)&$\num{6.0 +- 1.0}$\\
      $L'_\mathrm{CO}$($\si{K.km.s^{-1}.pc^2}$)&$\num{5.3+-0.9e+9}$\\
      $M_\mathrm{gas}$($M_\odot$)&$\num{2.3+-0.4e+10}$\\
      $\mu_\mathrm{gas}$&$\num{0.50+-0.22}$\\
      $t_\mathrm{depl}$($\si{Gyr}$)&$\num{2.9+-1.0}$\\
      $\mathrm{FWHM}$($\si{km.s^{-1}}$)$^{**}$&$\num{346+-49}$\\
      $S_\mathrm{1.3mm}$($\si{mJy}$)$^{*}$&$< \num{0.28}$\\
      \hline
      \uline{the western source}&\\
      $S_\mathrm{CO}\Delta V$($\si{Jy.km.s^{-1}}$)&$\num{7.5 +- 0.8}$\\
      $L'_\mathrm{CO}$($\si{K.km.s^{-1}.pc^2}$)&$\num{6.7+-0.7e+9}$\\
      $M_\mathrm{gas}$($M_\odot$)&$\num{2.9+-0.3e+10}$\\
      $t_\mathrm{depl}$($\si{Gyr}$)&$\num{1.4+-0.2}$\\
      $\mathrm{FWHM}$($\si{km.s^{-1}}$)$^{**}$&$\num{374+-29}$\\
      $S_\mathrm{1.3mm}$($\si{mJy}$)&$\num{0.17+-0.07}$\\
      \hline
    \end{tabular}
    \tablecomments{$^{*}$3$\sigma$ upper limit\,$^{**}$FWHM of CO(2--1) line width obtained by Gaussian fit.}
\end{table}

\subsection{HG\,20191001A}
\begin{figure*}[htbp]
    \centering
    \includegraphics[width=18cm]{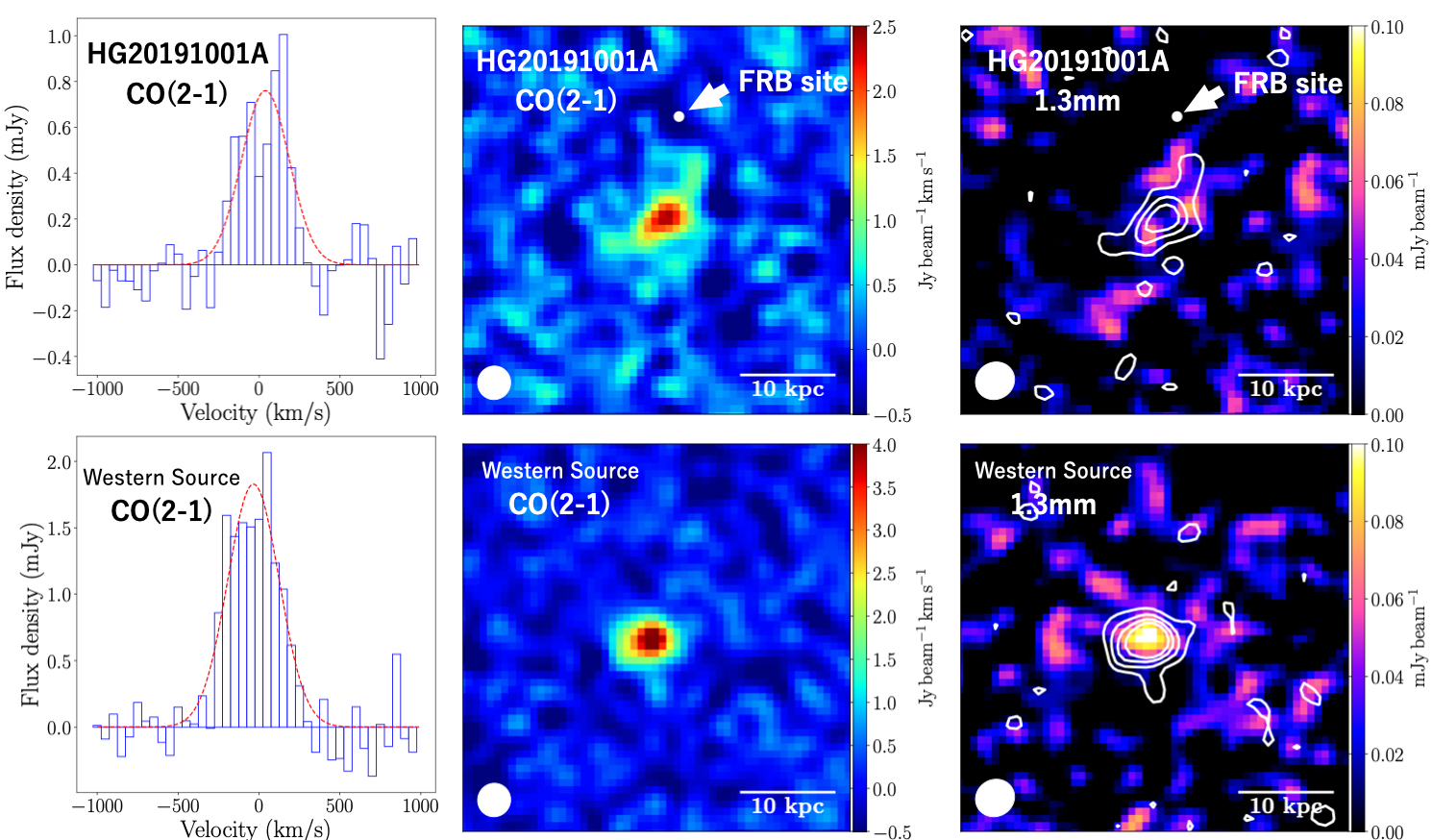}
    \caption{Results of the ALMA observations of HG\,20191001A (top panels) and the western source (bottom panels). The synthesized beam size is shown in the lower-left corner of the center and right maps. \textbf{Left.} CO(2--1) spectra. Velocity widths per bin are $\SI{50}{km.s^{-1}}$. The red dotted lines are fitted profiles with a Gaussian function. \textbf{Center.} CO(2--1) integrated intensity maps. \textbf{Right.} 1.3 mm continuum map overlaid with white contours of the CO(2--1) integrated intensities (the contour levels are the same as in Figure \ref{I-band}).}
    \label{obsresults}
\end{figure*}

The upper panels in Figure \ref{obsresults} show the CO(2--1) line profile, CO(2--1) integrated intensity map, and continuum map of HG\,20191001A, from left to right.
The CO line is spatially resolved, which is the first time for the FRB host galaxy in the cosmological distance. 
The peak signal-to-noise ratio in the cube is $\mathrm{S/N}=9.9$.

Using the Gaussian function to fit the CO line profile, the velocity width of the galaxy (at the Full Width at Half Maximum; FWHM) was estimated to be $\SI{346+-49}{km.s^{-1}}$.
For the 1.3 mm continuum, no $\mathrm{S/N}>3$ detection was found. 
Using the same region where the CO(2--1) spectrum was extracted, we derived the $3\sigma$ upper limit to the total continuum flux to be $\SI{0.28}{mJy}$.

CO luminosity is calculated from the following equation by \cite{Solomon2004}:
\begin{equation}
    L'_{\mathrm{CO(2-1)}}=\num{3.25e+7}\,S_\mathrm{CO}\Delta V\nu_\mathrm{obs}^{-2}D_L^2(1+z)^{-3}, 
\end{equation}
where $L'_{\mathrm{CO(2-1)}}$ is the line luminosity of CO(2--1) in units of $\si{K. km.s^{-1} .pc^{-2}}$, $S_\mathrm{CO(2-1)}\Delta V$ is the velocity-integrated flux in $\si{Jy. km.s^{-1}}$, $\nu_{\mathrm{obs}}$ is the observed frequency in GHz and $D_L$ is the luminosity distance in Mpc. 
In this paper, we adopt $\nu_{\mathrm{obs}} = \SI{186.822}{GHz}$, and $D_L=\SI{1179}{Mpc}$ that were calculated from optical redshift. 
The line luminosity is calculated as $L'_{\mathrm{CO(2-1)}}=\SI{5.3+-0.9e+9}{K .km.s^{-1} .pc^{-2}}$. 
The molecular gas mass $M_\mathrm{gas}$ of the galaxy is calculated from
\begin{equation}
    M_\mathrm{gas}=\alpha_{\mathrm{CO}}L'_\mathrm{CO(1-0)},
\end{equation}
where $\alpha_\mathrm{CO}$ is the CO-to-$\mathrm{H_2}$ conversion factor in units of $\si{M_{\odot}. (K.km.s^{-1}.pc^{-2})^{-1}}$, $L'_{\mathrm{CO(1-0)}}$ is the line luminosity of CO(1--0) in $\si{K. km.s^{-1} .pc^{-2}}$. 
$L'_\mathrm{CO(1-0)}$ was calculated assuming the typical star-forming galaxy value of $L'_\mathrm{CO(2-1)}/L'_\mathrm{CO(1-0)}=0.77$ \citep{Genzel2015,Tacconi}. 
The conversion factor $\alpha_\mathrm{CO}$ may depend on the gas-phase metallicity, but since the host's value $12+\log(\mathrm{O/H})=\num{8.9+-0.1}$ is similar to solar metallicity, we use the Galactic value $\alpha_\mathrm{CO}=\SI{4.3}{M_\odot. (K. km.s^{-1} pc^{-2})^{-1}}$. 
The validity of the conversion factor will be discussed in sec. \ref{sec:modeling}. 
The calculated host's gas mass is $M_\mathrm{gas}=\SI{2.3+-0.4e+10}{M_\odot}$.

\begin{figure}[htbp]
    \centering
    \includegraphics[width=9cm]{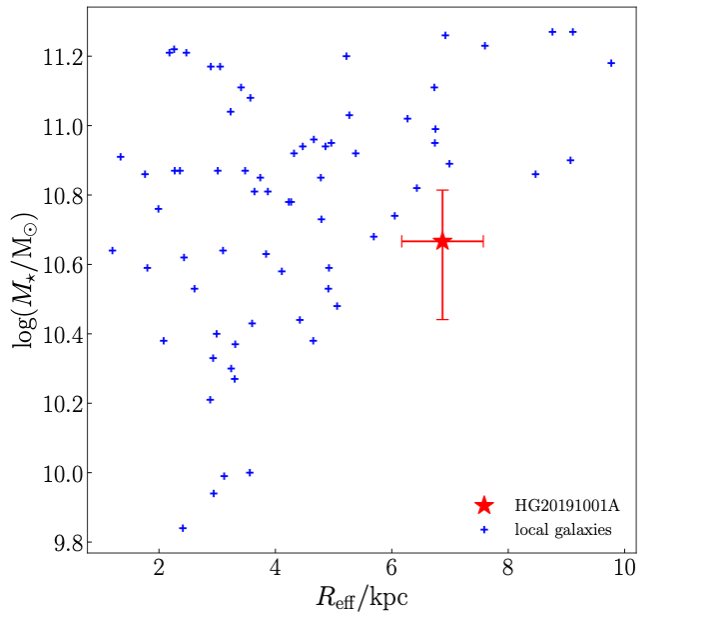}
    \caption{Stellar mass against CO half-light radius. The red star indicates HG\,20191001A, while blue crosses are for local star-forming galaxies from the EDGE CALIFA survey \citep{Bolatto2017}}
    \label{Reff}
\end{figure}

Figure \ref{Reff} compares the CO half-light radius of HG\,20191001A with that of nearby star-forming galaxies from the EDGE CALIFA survey \citep{Bolatto2017}. 
This figure shows that the CO disk of HG\,20191001A is slightly larger than that of star-forming galaxies with similar stellar mass.

\begin{figure*}[htbp]
    \centering
    \includegraphics[width=18cm]{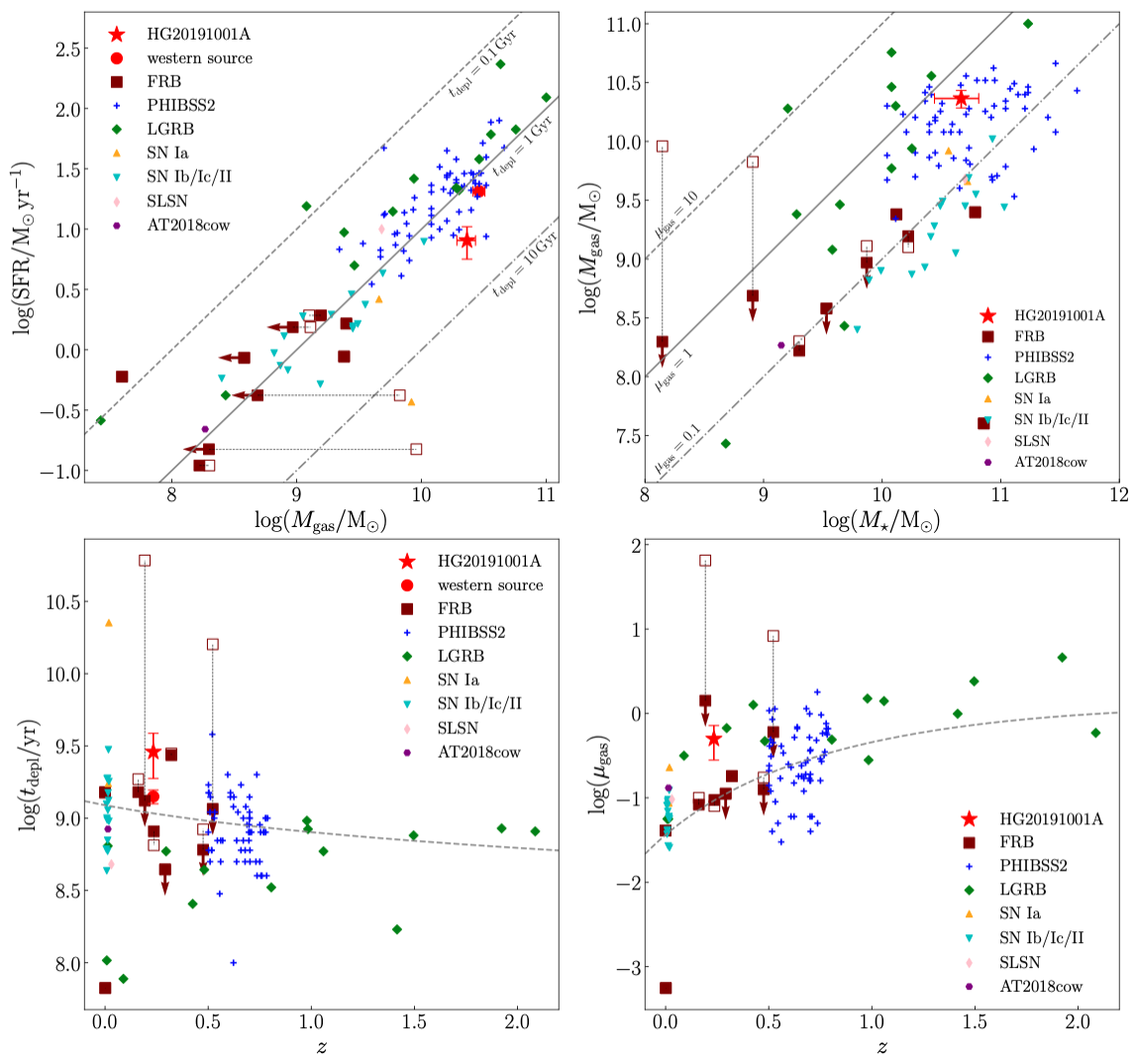}
    \caption{Comparison of molecular gas properties. For comparison, FRB hosts of previous studies \citep{Hatsukade2022, Chittidi2023}, star-forming galaxies from $z=0.2$ to 0.5 of the PHIBBS2 survey \citep{Freundlich2018}, long-duration GRB host galaxies \citep{Hatsukade2020}, host galaxies of SNe Ia, Ib/Ic/II \citep{Galbany2017}, super luminous supernova hosts \citep{Hatsukade2018}, and AT2018cow \citep{Morokuma-Matsui2019} are plotted. For FRB hosts, molecular gas mass and related values calculated with the Galactic conversion factor are presented by brown filled squares. Those with conversion factors in the original paper are marked with brown open squares. The upper limit objects are marked with arrow symbols. \textbf{Upper left.} Comparison of molecular gas mass and SFR. On the gray lines, gas depletion time is constant. \textbf{Upper right.} Stellar mass and gas mass. On the gray lines, the molecular gas fraction is constant. \textbf{Lower left.} Redshift and gas depletion time. 
    \textbf{Lower right.} Redshift and molecular gas fraction. The gray dashed lines in the lower panels indicate empirical relationships for the main sequence star-forming galaxies \citep{Tacconi}. }
    \label{gasprop}
\end{figure*}

In the $M_\mathrm{gas}-\mathrm{SFR}$ diagram in Figure \ref{gasprop}, HG\,20191001A is located at the position of the star mark. 
Compared to the previously observed FRB host galaxies (brown-filled squares from \cite{Hatsukade2022} and \citet{Chittidi2023}), HG\,20191001A is more gas-rich and actively star-forming.
The brown-filled squares represent the values calculated with the galactic conversion factor, and the open squares represent the values calculated with the conversion factor used in the original papers \citep{Hatsukade2022, Chittidi2023}. 
While a majority of the FRB hosts have lower molecular gas mass than the long-duration GRB hosts, some FRB hosts, including HG\,20191001A, are comparable to the GRB hosts, showing the diversity of FRB hosts.
Although the area occupied by FRB and SNe host galaxies \citep{Galbany2017} are similar, some outliers some outliers exist.
The sample size is insufficient to show that this difference is statistically significant, and further verification is expected.
Similarly in the $M_\mathrm{gas}-M_\star$ diagram, FRB host galaxies are located in a wide range on the diagram.
In the $t_\mathrm{depl}-z$ and $\mu_\mathrm{gas}-z$ diagrams (where $t_\mathrm{depl}=M_\mathrm{gas}/\mathrm{SFR},\,\mu_\mathrm{gas}=M_\mathrm{gas}/M_\star$), HG\,20191001A appears to be located not far from the empirical trend line of star-forming galaxies \citep[gray dashed line;][]{Tacconi} and the hosts of other transients.
The FRB host galaxies do not show any clear redshift dependence but may be widely distributed along the vertical axis, depending on the conversion factor used.

\subsection{the western source}\label{sec:res_western}
The bottom three figures in Figure \ref{obsresults} show the line profile, CO integrated intensity map, and continuum map for the western source.
The peak signal-to-noise ratio in the cube is $\mathrm{S/N}=13$ and the FWHM of the velocity width is $\SI{374+-29}{km.s^{-1}}$.
The 1.3 mm continuum was detected with $\mathrm{S/N}=3.6$ and the continuum flux was $S_\mathrm{1.3mm}=\SI{0.17+-0.07}{mJy}$.
Using the same assumptions as in the previous section, $L'_\mathrm{CO(2-1)}=\SI{6.7+-0.7e+9}{K.km.s^{-1}.pc^{-2}}$ and $M_\mathrm{gas}=\SI{2.9+-0.3e+10}{M_\odot}$ were calculated.
The gas mass for the western source is slightly larger than that for the host.

\subsection{FRB\,20191001A site}
FRB\,20191001A was detected 11 kpc north of the center of the host, and no CO was detected within the FRB localization uncertainty  \citep[$\sigma_R=0\farcs149$;][]{Heintz2020}. 
The 3$\sigma$ upper limit for the molecular gas column density ($N(\mathrm{H_2})$) is $< \SI{2.1e+21}{cm^{-2}}$.
That is obtained by assuming the Galactic conversion factor, and the CO linewidth at the site, $\SI{100}{km.s^{-1}}$, which is the velocity width of the edge part of the galaxy (see Section  \ref{sec:modeling}).

Some theoretical models predict that millimeter emission is enhanced after a few years of the FRB event \citep[e.g.,][]{Murase2016, Margalit2018}, but no millimeter emission has been detected in previous observations \citep[e.g.,][]{Eftekhari2021}. 
No 1.3 mm continuum was detected at the FRB site in this study
and the 3$\sigma$ upper limit 1090 days after the FRB detection is $< \SI{9.1e-2}{mJy.beam^{-1}}$. 

\begin{figure*}[htbp]
    \centering
    \includegraphics[width=18cm]{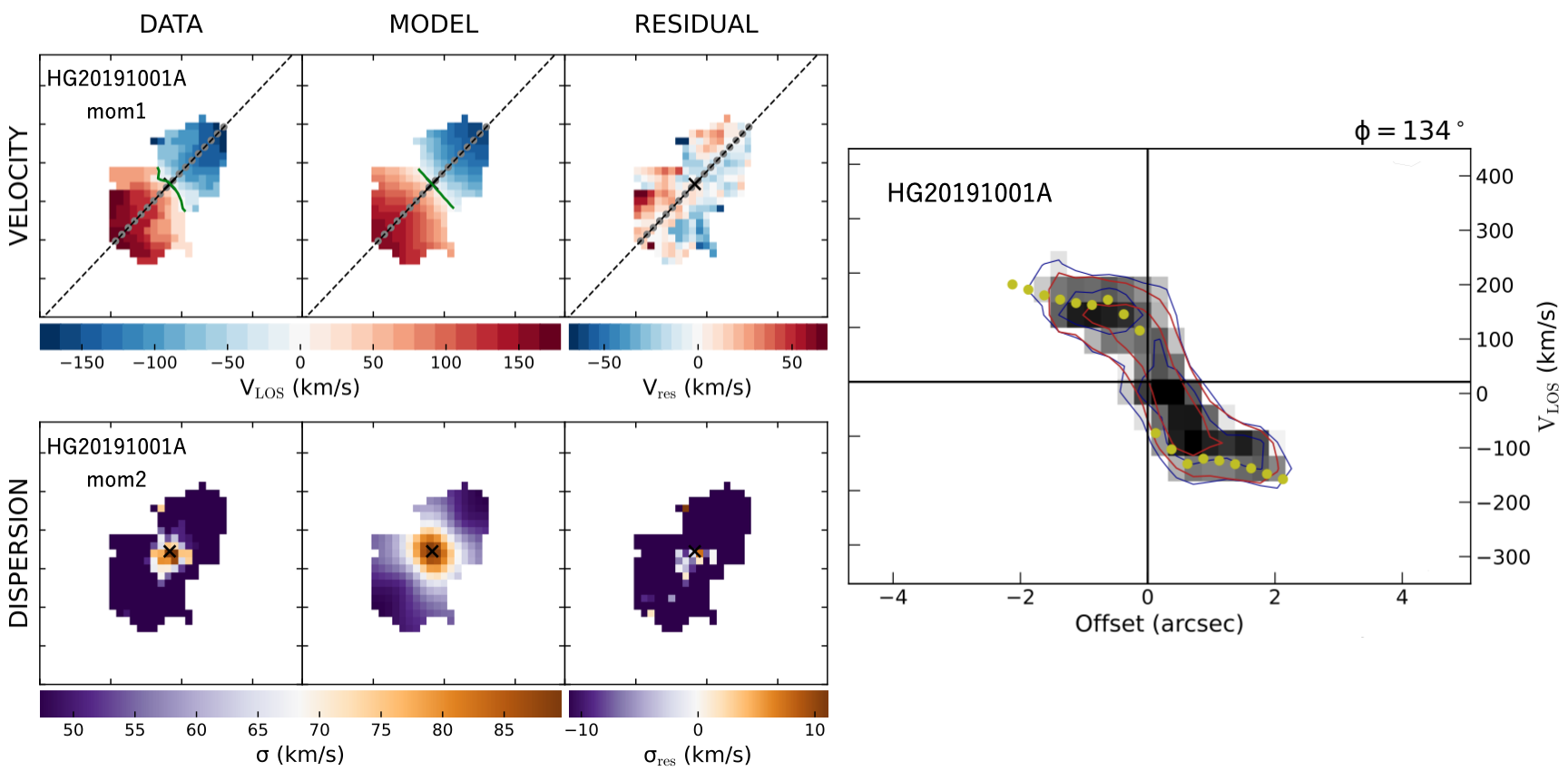}
    \caption{The results of the $^\mathrm{3D}$Barolo modeling of the CO(2--1) emission in HG\,20191001A. \textbf{Left.} Observed maps, modeled maps, and residuals of the velocity field and the velocity dispersion. We note that the velocity dispersion includes instrumental broadening of $\SI{50}{km.s^{-1}}$. \textbf{Right.} Observed (gray scale) and modeled (contours) position-to-velocity diagrams along the major axis (134$^\circ$), which is shown in the left panels as diagonal dashed lines. The derived rotation velocities are also shown in yellow dots. }
    \label{modeling}
\end{figure*}

\begin{figure*}[htbp]
    \centering
    \includegraphics[width=18cm]{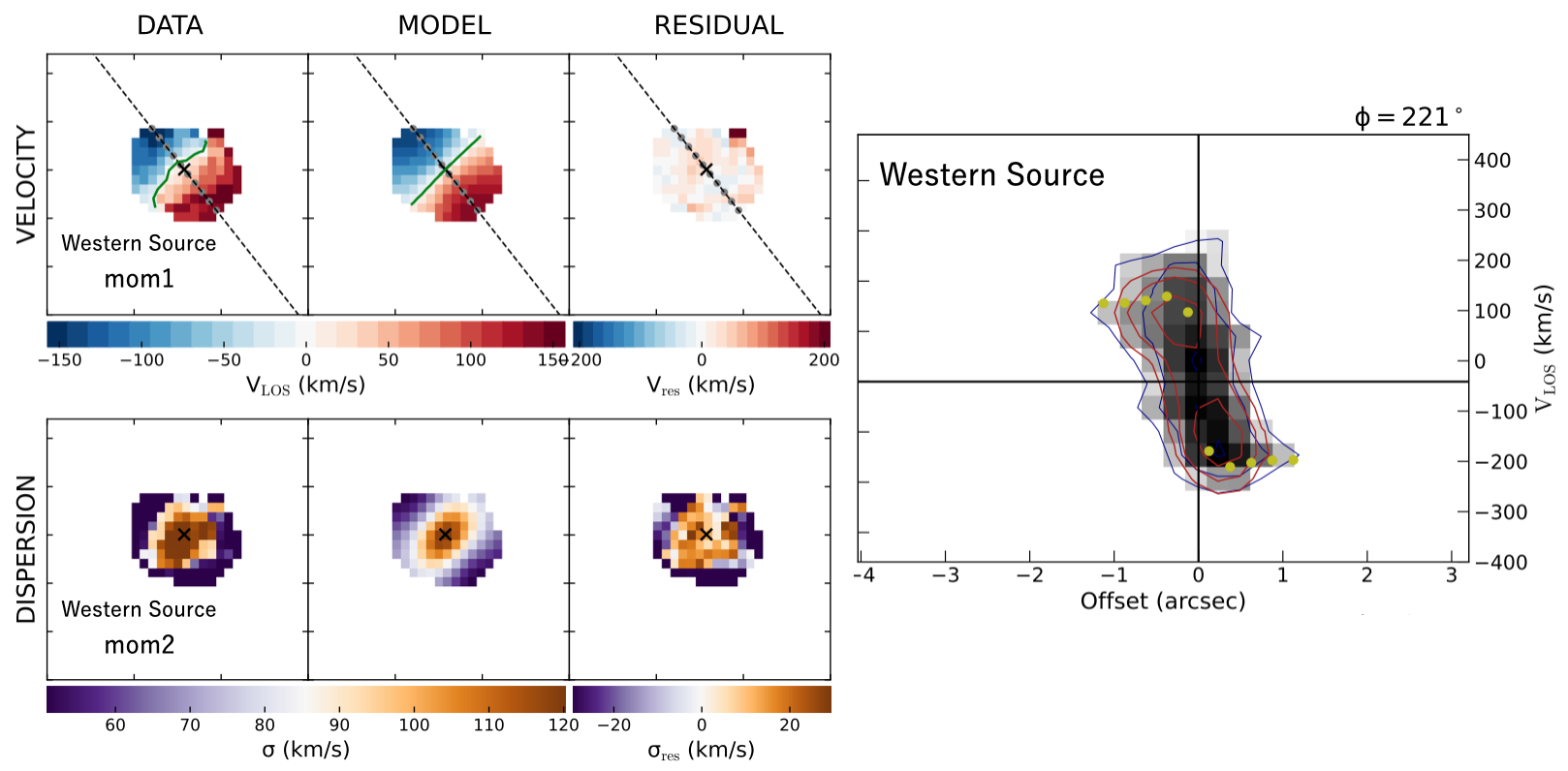}
    \caption{Same as Figure \ref{modeling} but for the western source ($\mathrm{P.A.=221^\circ}$).}
    \label{modeling_w}
\end{figure*}

\subsection{Modelling CO disk dynamics}\label{sec:modeling}
CO disk dynamics of both galaxies were modeled by $^\mathrm{3D}$Barolo \citep{Teodoro2015}.
$^\mathrm{3D}$Barolo fits tilted-ring models to a data cube.
The ring width was set to $0\farcs25$ and modeling was performed to the CO detection limit, and the normalization option was set to local.
In the first estimation, rotational velocity, velocity dispersion, systematic velocity, inclination, and position angle were set to free, and in the second estimation, the values of inclination and position angle were fixed between rings for modeling.
The parameter VRAD, the galactocentric radial velocity, was fixed to 0 in the modeling, assuming no radial motion of the molecular gas.
The results of the modeling are shown in Figures \ref{modeling} and \ref{modeling_w} and Table \ref{modeling_val}.
The small velocity dispersion ($V_\mathrm{disp}$ in Table \ref{modeling_val}) and the position velocity (PV) diagrams (Figures \ref{modeling} and \ref{modeling_w}) indicate that molecular gas disks of both galaxies are rotation dominated. 
The ratio of the rotational velocity to the velocity dispersion is 49 and 42 for HG\,20191001A and western source, respectively. These values are higher than 1, supporting the rotation-dominated systems \citep{Pelliccia2017}.
We do not find a clear sign of interaction between these galaxies.
The difference between the derived CO systemic velocity and the optical systemic velocity is small for the host.
We fit an exponential function to the radial profile of surface brightness and derive the half-light radius to be $\SI{6.9+-0.7}{ kpc}$ for the host and $\SI{3.0+-0.5}{kpc}$ for the western source.
We derive the dynamical mass $M_\mathrm{dyn}$ using the rotational velocity $V_\mathrm{rot}$ as follows:
\begin{equation}
    M_\mathrm{dyn}=\frac{RV_\mathrm{rot}^2}{G}.
\end{equation}
The dynamical mass at $R=\SI{8.0}{kpc}$, the CO detection limit, is $M_\mathrm{dyn}=\SI{7.8 +- 2.0e+10}{M_\odot}$ for the host.
For the western source, $R=\SI{4.2}{kpc}$ and $M_\mathrm{dyn}=\SI{1.3 +- 0.4e+11}{M_\odot}$.
For the host, since the sum of the stellar mass and gas mass is smaller than the dynamical mass, an upper limit for $\alpha_\mathrm{CO}$ can be determined as $\alpha_\mathrm{CO} < \SI{5.9}{M_\odot.(K .km.s^{-1}. pc^{-2})^{-1}}$. 
This result is consistent with the value of $\alpha_\mathrm{CO}$ assumed in this study.

\begin{table}[htbp]
  \caption{Results of the $^\mathrm{3D}$Barolo modeling}
  \label{modeling_val}
  \centering
    \begin{tabular}{lcc}
      \hline \hline
      Parameter&HG\,20191001A&Western source\\
      \hline
      $R$($\mathrm{kpc}$)$^{*}$&8.0&4.2\\
      $R_\mathrm{eff}$(kpc)&$\num{6.9+-0.7}$&$\num{3.0+-0.5}$\\
      $i$(deg)&$\num{61+-5}$&$\num{25+-4}$\\
      P.A.(deg)&$\num{134+-6}$&$\num{221+-8}$\\
      $V_\mathrm{rot}$($\si{km.s^{-1}}$)&$\num{205+-26}$&$\num{366+-54}$\\
      $V_\mathrm{sys}$($\si{km.s^{-1}}$)$^{**}$&$\num{21+-7}$&$\num{-41+-8}$\\
      $V_\mathrm{disp}$($\si{km.s^{-1}}$)$^{***}$&$\num{4.2+-12.5}$&$\num{8.7+-19.1}$\\
      $M_\mathrm{dyn}$($M_\odot$)&$\num{7.8+-2.0e+10}$&$\num{1.3+-0.4e+11}$\\
      \hline
    \end{tabular}
    \tablecomments{$^{*}$Radius of CO detection limit.\,$^{**}$Systematic velocity where the galaxy's restframe velocity calculated from optical redshift is defined as $\SI{0}{ km.s^{-1}}$ \,$^{***}$Mean value of velocity dispersion at each ring, but the values at the inner three rings were removed because of the significant influence of the initial values in the modeling. The velocity dispersion is deconvolved with the spectral resolution of $\SI{50}{km.s^{-1}}$. }
\end{table}

\section{Discussion}
\subsection{Molecular gas properties of the host galaxy}
\cite{Hatsukade2022} argued that the gas mass and SFR of the FRB host galaxies take a wide range of values and suggested that they may not originate from a single source, which depends on the molecular gas properties, or they may originate from multiple sources that can take a wide range of values. 
Our observations show that the range of molecular gas property values, especially $M_\mathrm{gas}$, of FRB host galaxies, has been expanded compared to previous observations.
This suggests that the scenario of a single source originating from a star formation-dependent environment, such as a massive star, would be disfavoured, whereas multiple sources or a source independent of molecular gas properties, such as old star populations, would be favoured as the FRB progenitors.
In this discussion, we do not distinguish between repeating and one-off FRBs due to the small sample size.
If the two origins are different, as is often argued \citep[e.g.,][]{Ravi2019, Hashimoto2020}, this is consistent with the ``multiple origin'' theory suggested by the molecular gas results.
To discuss the two populations separately, a larger sample is necessary.

\subsection{Molecular gas property at the FRB site}
\begin{figure}
    \centering
    \includegraphics[width=9cm]{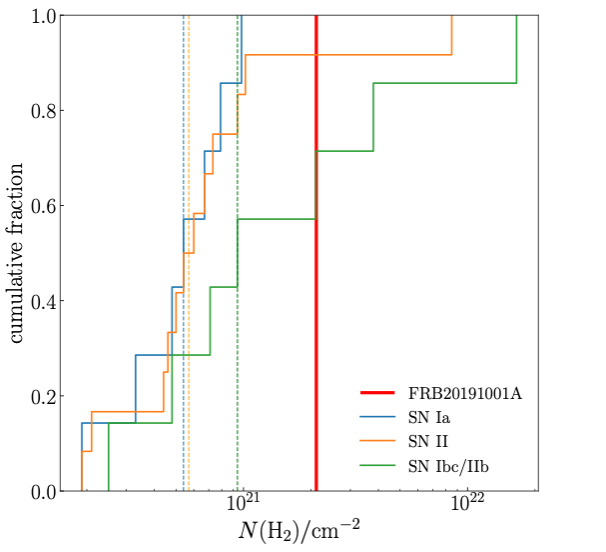}
    \caption{3$\sigma$-upper limit of molecular gas column density at the site of FRB\,20191001A. For comparison, cumulative fraction of supernova type Ia, II, Ibc/IIb \citep{Galbany2017} are presented. The median values of each type of SN are also shown in dotted vertical lines.}
    \label{NH2}
\end{figure}

While the FRB environments can be estimated from molecular gas properties of their host galaxies, a more direct discussion would benefit from examining the molecular gas at the site where the FRBs are detected.
\cite{Galbany2017} conducted CO observations of the host galaxies of SN type Ia, Ibc/IIb, and II, and derived the molecular gas column density at those sites. 
Figure \ref{NH2} shows the 3$\sigma$ upper limit of the molecular gas column density ($N(\mathrm{H_2})$) for FRB\,20191001A and compares it with the cumulative distribution of the SNe \citep{Galbany2017}. 
The upper limit (the red solid line) is larger than the median value of the gas column density for each type of SN (dashed lines), and thus does not constrain its progenitor.
The limiting factor here is the non-detection of CO at the FRB site, and deeper observations may add a limit to the source of progenitor.
We would like to note that this study is the first to discuss the molecular gas properties at an FRB site. 
More detection will allow us to create a cumulative distribution and thereby make statistical comparisons with other possible progenitors. 
Further observations with high spatial resolution and sensitivity of FRB hosts and at those sites are needed for this purpose.

\subsection{Galactic interaction}
HI observations of FRB host galaxies which show features of interaction have been reported by \cite{Kaur2022}, \cite{Micha2021}, and \cite{Glowacki2023}. 
\cite{Hsu2022} shows the disturbed kinematic structure in the FRB host from the CO observation.
HG\,20191001A is thought to interact with the western source, and we discuss the extent of the interaction and its effect on FRB generation. 
We cannot see asymmetric or disrupted structures in the modeled velocity fields and the position-to-velocity diagrams (see Figures \ref{modeling},\ref{modeling_w}).
That means that the velocity fields of the host and the western source are well-fitted with normal rotation, and the effect of the interaction between them can be weak in terms of molecular gas kinematics.
The impact of the interaction can also be investigated from star formation activity.
For example, \cite{Pan2018} showed that the SFR and gas mass are enhanced in galaxies in pairs.
In the $M_\mathrm{gas} - \mathrm{SFR}$ diagram (Figure \ref{gasprop}), the host and the western source are located near star forming galaxies at $z=0.2$--$0.5$ from the PHIBSS2 survey \citep{Freundlich2018}. This result indicates that both galaxies are not experiencing starburst.
\cite{Kaur2022} proposed a scenario in which gases are dynamically disrupted by galactic interactions and the resulting star formation activation is attributed to FRB generation, but the above discussion suggests that FRB\,20191001A is not likely to follow that scenario.

\section{SUMMARY}
We observed the CO(2--1) emission of the host galaxy of FRB\,20191001A (HG\,20191001A) along with its close ($\sim \SI{25}{kpc}$) companion galaxy (the western source) using ALMA and obtained the following results and findings.
\begin{itemize}
    \item The CO(2--1) emission lines of HG\,20191001A and its companion, the western source, were successfully detected, and are spatially resolved for HG\,20191001A, for the first time for an FRB host at cosmological distances. This FRB host galaxy has a larger molecular gas mass than other known FRB hosts with CO observations. Its disk size is also relatively large compared to nearby star-forming galaxies.
    \item The measured molecular gas mass of HG\,20191001A, $\SI{2.3+-0.4e+10}{M_\odot}$, has expanded the parameter space for molecular gas mass in FRB host galaxies, and that strengthening the possibility that FRBs originate from multiple sources or sources unrelated to recent star formation. 
    \item We discussed local molecular gas environments of an FRB for the first time. The obtained 3$\sigma$ upper limit of molecular gas column density at the site of FRB\,20191001A is $\SI{2.1e+21}{cm^{-2}}$. Deeper observations are needed to constrain its progenitor.
    \item We performed dynamical modeling of the CO disk of HG\,20191001A and the western source and found that they are both rotation-dominated. The effect of the interaction of HG\,20191001A with the western source, both in terms of kinematics and star formation, is unlikely to be strong. A larger sample is needed to further discuss whether galactic interactions trigger FRBs.
\end{itemize}

\begin{acknowledgments}
We would like to thank Kasper E.\ Heintz for sharing his $I$-band image of HG\,20191001A.
This work is supported by JSPS KAKENHI grant No.\ 19K03925 and 23K03449 (B.H.), 
17K14259 (F.E.), 20H00172 (F.E., H.K.), 17H06130 (K.K.), 
and the NAOJ ALMA Scientific Research Grant Number 2020-15A (H.K.). 
T.H.\ acknowledges the support of the National Science and Technology Council of Taiwan through grants 110-2112-M-005-013-MY3, 110-2112-M-007-034-, and 111-2123-M-001-008-.
This paper makes use of the following ALMA data: ADS/JAO.ALMA\#2021.1.00027.S. ALMA is a partnership of ESO (representing its member states), NSF (USA) and NINS (Japan), together with NRC (Canada), MOST and ASIAA (Taiwan), and KASI (Republic of Korea), in cooperation with the Republic of Chile. The Joint ALMA Observatory is operated by ESO, AUI/NRAO and NAOJ.
Data analysis was carried out on the Multi-wavelength Data Analysis System operated by the Astronomy Data Center (ADC), National Astronomical Observatory of Japan.
\end{acknowledgments}

\bibliography{FRB191001}
\bibliographystyle{aasjournal}
\end{document}